\documentclass[twocolumn,aps,pra,10pt,showpacs]{revtex4-2}
\usepackage{bm}
\usepackage{amsfonts}
\usepackage{amssymb}
\usepackage{amsmath}
\usepackage{graphicx}

\begin{document}

{\bf Comment on "Fractional Angular Momenta, Gouy and Berry Phases in Relativistic Bateman-Hillion-Gaussian Beams of Electrons"}\\

In the recent Letter \cite{dph} the authors derived Dirac bispinor wave functions of the Laguerre-Gauss type using the extension of the Bateman-Hillion construction originally applied to the d'Alembert equation. The purpose of this Comment is to show that the Laguerre-Gauss solutions (and many more) can be easily obtained by simple operations on generating functions. One generating function leads to the whole family of solutions. I will discuss here the solutions of the Klein-Gordon equation since the solutions of the Dirac equation can be obtained \cite{bb} by differentiation.

This generating function for the Legendre-Gauss solutions of the Klein-Gordon equation is a simple Gaussian function,
\begin{align}\label{gen}
G_{\rm LG}(x_+,x_-,z,t)&=\exp\left(-i E t_-/2\right)\exp\left(-i\frac{m^2}{2E}t_+\right)
\nonumber\\&\times\frac{1}{a(t_+)}
\exp\left(-\frac{x_+x_-}{a(t_+)}\right),
\end{align}
where $x_\pm=x\pm iy$, $t_\pm=t\pm z/c$, $a(t_+)=w_0^2+2it_+/E$, and $\hbar=1,\,c=1$. The parameter $w_0$ is a measure of the beam waist. The generation of the LG beams by differentiation requires a slight extension of the classic Rodrigues formula \cite{gr} for the Laguerre polynomials,
\begin{align}\label{lg0}
f_{\rm LG}^{nl}&(x,y,z,t)=\frac{(-1)^{n+l}}{n!}
\frac{\partial^{n+l}}{\partial x_-^{n+l}}\frac{\partial^n}{\partial x_+^n}G_{\rm LG}(x,y,z,t).
\end{align}
The resulting expressions are \cite{bb},
\begin{align}\label{lg}
f_{\rm LG}^{nl}(x,y,z,t)=\exp\left(-i E t_-/2\right)\exp\left(-i\frac{m^2}{2E}t_+\right)
\nonumber\\\times\frac{(x+iy)^{l}}{a(t_+)^{n+l+1}}
\exp\left(-\frac{x^2+y^2}{a(t_+)}\right)
L_n^{l}\left(\frac{x^2+y^2}{a(t_+)}\right).
\end{align}
They differ from the formulas given in \cite{dph} since the argument of the Laguerre polynomials in (\ref{lg}) involves the Gouy phase, whereas in \cite{dph} this argument is real. This difference is due to a different meaning of the quantum number $n$ in (\ref{lg}) and the quantum number $p$ appearing in Eq.~(5) of \cite{dph}. The function (\ref{lg}) for a given $n$ turns out to be a linear combination of the functions in \cite{dph} with $p=n,\,p=n-1,\,{\rm etc.}$

The same Gaussian generating function $G_{\rm HG}(x,y,z,t)=G_{\rm LG}(x_+,x_-,z,t)$, but treated as a function of $x$ and $y$, can be used to obtain the Hermite-Gauss beams. This time one has to use the Rodrigues formula for Hermite polynomials and evaluate the derivatives with respect to $x$ and $y$,
\begin{align}\label{h0}
&f_{\rm HG}^{mn}(x,y,z,t)=(-1)^{m+n}\frac{\partial^m}{\partial x^m}\frac{\partial^n}{\partial y^n}G_{\rm HG}(x,y,z,t)\nonumber\\
&=H_m\left(\frac{x}{\sqrt{a(t_+)}}\right)
H_n\left(\frac{y}{\sqrt{a(t_+)}}\right)
\frac{G_{\rm HG}(x,y,z,t)}{a(t_+)^{(m+n)/2}}.
\end{align}
These functions satisfy the KG equation but they again differ  from the functions defined by the same authors in \cite{dp}, by having complex arguments of the Hermite polynomials.

The list of generating functions does not end with Laguerre-Gauss and Hermite-Gauss. For example, choosing the generating function in the following exponential form,
\begin{align}\label{exp}
G_{\rm Exp}&=e^{iq z}e^{-u}/u,\\
u&=\sqrt{(w_0+i\sqrt{m^2+q^2}t)^2+(m^2+q^2)x_+x_-},
\end{align}
one obtains by differentiation with respect to $x_-$ a family of beam-like solutions of the KG equation. Like the LG beams, they carry angular momentum and their waist is determined by $w_0$ but this time they do not have the Gaussian but the exponential fall-off in the transverse direction. Another example of a generating function with an exponential fall-off is the Macdonald function \cite{bb1},
\begin{align}\label{md}
G_{\rm Md}&=m K_1(m s)/s,\\
s&=\sqrt{(w_0+it)^2+x_+x_-+z^2}.
\end{align}

The generating functions that in addition depend on some parameters allow for the construction of new solutions in a diffferent way. One may not only differentiate them with respect to coordinates (as in (\ref{lg0}) and (\ref{h0}) for Laguerre-Gauss and Hermit-Gauss beams) but one may also obtain new solutions by differentiation and integration with respect to these parameters. For example, the standard Bessel solutions of the KG equations are obtained from the generating function $G_{\rm B}$,
\begin{align}\label{genb}
G_{\rm B}(\rho,\phi,z,t)&=\exp\left(-i\sqrt{p_\perp^2+p_z^2+m^2} t+ip_z z\right)\nonumber\\
&\times\exp\left(ip_\perp\rho\cos(\phi-\varphi)\right),
\end{align}
by integrating the product $e^{il\varphi}G_{\rm B}$ over $\varphi$ from 0 to $2\pi$.

As a similar but less trivial example, let us consider the following solution of the KG equation,
\begin{align}\label{genbg}
&G_{\rm BG}(\rho,\phi,z,t)=\exp\left(-i E t_-/2\right)\exp\left(-i\frac{m^2}{2E}t_+\right)
\nonumber\\&\times\frac{1}{a(t_+)}
\exp\left(-\frac{\rho^2-b^2+2ib\rho\cos(\phi-\varphi)}
{a(t_+)}\right),
\end{align}
where $b$ and $\varphi$ are arbitrary parameters. Again, by integrating this function multiplied by $e^{il\varphi}$ from $0$ to $2\pi$ with respect to the parameter $\varphi$ one obtains a new family of solutions of the KG equation,
\begin{align}\label{bg}
&f_{\rm BG}^{l}(\rho,\phi,z,t)=2\pi(-i)^l\frac{e^{il\phi}}
{a(t_+)}\exp\left(-i E t_-/2\right)\nonumber\\
&\times\exp\left(-i\frac{m^2}{2E}t_+\right)
\exp\left(-\frac{\rho^2-b^2}{a(t_+)}\right)
J_l\left(\frac{2b\rho}{a(t_+)}\right),
\end{align}
which may be called Bessel-Gauss solutions. The advantage of using the generating function is clearly seen because it is much easier to check that the function (\ref{genbg}) satisfies the KG equation than to do this for the Bessel beams (\ref{bg}).

The generating functions for the KG equation in the limit $m\to 0$ become solutions of the d'Alembert equation. From these solutions with the use of the Whittaker construction one obtains solutions of Maxwell equation, as explained in \cite{rs}. This fact leads to the intriguing one-to-one correspondence between solutions of the Dirac equation obtained from the generating functions and the solutions of the Maxwell equations.

In conclusion, I have shown that there is an alternative to using the fairly involved Bateman-Hillion construction to obtain the Lagerre-Gauss and Hermite-Gauss beams. These beams, and many more, may be obtained from the generating functions. This method is easy to control because one must only make sure that the {\em generating function} obeys the KG equation.
\vspace{0.4cm}

\noindent Iwo Bialynicki-Birula\\
Center for Theoretical Physics\\
Polish Academy of Sciences\\
Aleja Lotnik\'ow 32/46, 02-668 Warsaw, Poland

\end{document}